%% file: colm2026_conference.tex
\documentclass{article} 
\usepackage[preprint]{colm2026_conference}

\usepackage{microtype}
\usepackage{hyperref}
\usepackage{url}
\usepackage{booktabs}

\usepackage{lineno}

\usepackage{graphicx}

\definecolor{darkblue}{rgb}{0, 0, 0.5}
\hypersetup{colorlinks=true, citecolor=darkblue, linkcolor=darkblue, urlcolor=darkblue}

\usepackage{amsmath}
\usepackage{amssymb}
\usepackage{mathtools}
\usepackage{amsthm}

\usepackage{makecell}
\usepackage{pifont} 
\usepackage[most]{tcolorbox}

\usepackage{microtype}
\usepackage{graphicx}
\usepackage{subcaption}
\usepackage{booktabs} 
\usepackage{longtable} 

\usepackage{xcolor}
\usepackage{amssymb}

\definecolor{darkgreen}{RGB}{0,100,0}
\definecolor{darkred}{RGB}{139,0,0}
\definecolor{swenextblue}{RGB}{78,121,167}
\definecolor{swenextbluebg}{RGB}{243,247,252}

\newcommand{\cmark}{\textcolor{darkgreen}{\checkmark}}
\newcommand{\xmark}{\textcolor{darkred}{\ensuremath{\times}}}

\newtcolorbox{swenextbox}[2][]{%
  enhanced,
  breakable,
  colback=swenextbluebg,
  colframe=swenextblue!75!black,
  coltitle=black,
  fonttitle=\bfseries,
  boxrule=0.8pt,
  arc=2mm,
  left=1.4mm,
  right=1.4mm,
  top=1.0mm,
  bottom=1.0mm,
  title={#2},
  #1
}

\newcommand{\SwenextCandidatesExecuted}{102582}
\newcommand{\SwenextFinalInstances}{2308}
\newcommand{\SwenextFinalRepos}{311}
\newcommand{\SwenextFinalYieldPct}{2.2}
\newcommand{\SwenextRepoQuarterProfiles}{1273}
\newcommand{\SwenextUniqueEnvSignatures}{173}
\newcommand{\SwenextCommitDateMin}{2012-01-31}
\newcommand{\SwenextCommitDateMax}{2025-12-25}
\newcommand{\SwenextTrajTotal}{8454}
\newcommand{\SwenextTrajSuccess}{3693}
\newcommand{\SwenextTrajSuccessPct}{72.6}
\newcommand{\SwenextTrajRanTestPct}{97.6}
\newcommand{\SwenextTrajRanReproPct}{90.3}

\theoremstyle{plain}

\theoremstyle{definition}

\theoremstyle{remark}

\title{SWE-Next: Scalable Real-World Software Engineering Tasks for Agents}

\author{
\hspace*{\leftmargini}%
\begin{minipage}[t]{\dimexpr\textwidth-2\leftmargini\relax}
\centering
{\bfseries Jiarong Liang$^{1*}$ \quad
Zhiheng Lyu$^{1*}$ \quad
Zijie Liu$^{2}$ \quad
Xiangchao Chen$^{1}$} \\
{\bfseries Ping Nie$^{3}$ \quad
Kai Zou$^{4}$ \quad
Wenhu Chen$^{1\dagger}$} \\[0.5em]
{\normalfont $^{1}$University of Waterloo \quad
$^{2}$University of North Carolina at Chapel Hill} \\
{\normalfont $^{3}$Independent \quad
$^{4}$Netmind.ai} \\[0.5em]
{\normalfont \url{https://github.com/TIGER-AI-Lab/SWE-Next}}
\end{minipage}
}

\begin{document}

\ifcolmsubmission
\linenumbers
\fi

\maketitle
\begingroup
\renewcommand{\thefootnote}{}
\footnotetext{$^{*}$Equal contribution \qquad $^{\dagger}$Corresponding author}
\endgroup

\begin{abstract}
Executable software engineering data is valuable for training SWE agents, but scaling it remains difficult for two reasons: only a small fraction of real repository changes yield verifiable, high-signal task instances, and naively building repository-specific environments quickly becomes the dominant systems cost. We present \textbf{SWE-Next}, an execution-grounded framework for scalable SWE task and trajectory collection. On the data side, SWE-Next mines real merged pull requests, executes candidate base/merged commit pairs, and retains only those that produce strict test improvements without regressions, yielding self-verifying instances. It also applies strict submission gating so that collected trajectories remain evidence-driven rather than speculative. On the systems side, SWE-Next introduces reusable \emph{repo-quarter profiles}, which reuse the same environment across nearby commits in time while keeping each task run separate and reproducible. Using only \textbf{30 hours} and \textbf{639\,GB} of environment storage, SWE-Next processes 3{,}971 seed repositories and 102{,}582 candidate commit pairs mined from real merged PRs to construct a dataset of 2{,}308 self-verifying instances. Experiments show that SWE-Next improves downstream pass@1 with fewer or comparable training trajectories, indicating that its gains come not from a stronger trajectory generator, but from higher-signal execution-grounded supervision and more efficient data collection.
\end{abstract}

\section{Introduction}

Large language model (LLM)-based agents are now used for concrete, multi-step tasks, such as web-scale retrieval and reasoning systems \citep{wei2025browsecompsimplechallengingbenchmark}, software engineering (SWE) agents evaluated on benchmarks \citep{jimenez2024swebench}, and GUI agents in realistic OS environments \citep{OSWorld}. Across these settings, effective agent development benefits from: tasks must be verifiable through executable environments, and they must be scalable via automated task synthesis and environment reuse.

Recent work has made important progress on executable SWE data collection, including task mining, environment construction, and test-based verification \citep{badertdinov2025swerebenchautomatedpipelinetask,yang2025swesmith,hu2025repo2run,jain2025r2egymproceduralenvironmentshybrid}. However, two challenges remain unresolved in combination: ensuring that collected tasks and trajectories are genuinely high-signal, and making large-scale environment management efficient enough for broad practical use.

\begin{figure*}[t]
  \centering
  \includegraphics[width=1\textwidth]{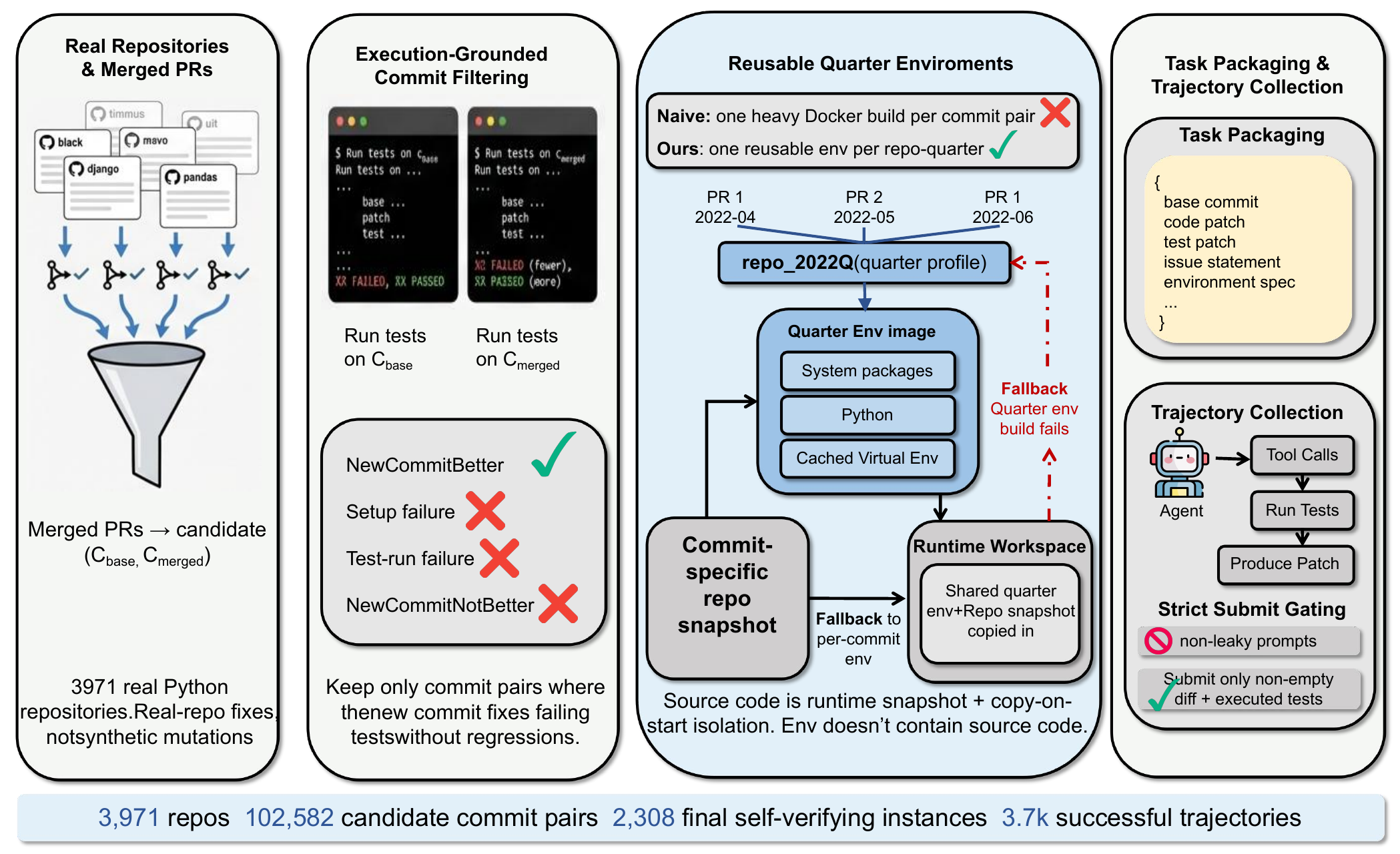}
  \caption{End-to-end overview of SWE-Next. Starting from real repositories and merged pull requests, we execute tests on base/merged commit pairs to retain verifiable instances, amortize environment setup via reusable quarter profiles, and package validated tasks for trajectory collection and post-training.}
  \label{fig:teaser}
\end{figure*}

In practice, three bottlenecks still limit the use of executable SWE tasks for post-training. First, although real repositories contain many candidate fixes, only a small fraction yield verifiable task instances and high-signal trajectories suitable for learning. Weak filtering, leaked supervision, or speculative submissions can all reduce training value. Second, environment configuration and execution still require substantial manual effort or ad hoc scripting. Researchers must resolve dependencies, verify that environments build and run correctly, handle failures, and connect these steps to task synthesis and verification. In many systems, environment management is also tied to cluster orchestrators such as Kubernetes. While effective for long-running production services, such infrastructure is unnecessarily heavy for research workloads that must repeatedly spin up and tear down short-lived, repository-specific environments. Third, executable task collection remains expensive at scale: without environment reuse, building and storing large numbers of repository-specific environments quickly becomes the dominant cost in large-scale data collection, motivating reusable environment profiles.

We address these challenges with \textbf{SWE-Next}, an execution-grounded framework for scalable SWE task and trajectory collection from real repositories. On the data side, SWE-Next executes candidate base/merged commit pairs and retains only those that produce strict improvements without regressions, yielding self-verifying task instances by construction. It also applies strict submission gating and non-leaky prompting so that collected trajectories remain evidence-driven rather than speculative. On the systems side, SWE-Next introduces reusable \emph{repo-quarter profiles}, which amortize environment construction and storage across temporally nearby commits, making large-scale collection substantially faster and cheaper.

This paper makes two main contributions.
(1) We present \textbf{SWE-Next}, an execution-grounded framework that turns real merged-PR commits into self-verifying SWE tasks, and pairs them with high-signal trajectory collection defaults such as strict submission gating. Figure~\ref{fig:teaser} provides an end-to-end overview.
(2) We introduce \textbf{repo-quarter profiles}, a reusable environment mechanism that amortizes build and storage cost across temporally nearby commits, substantially reducing resource requirements and accelerating large-scale executable SWE data collection.

In the run described in this paper, SWE-Next processes 3,971 repositories spanning diverse software domains, executes 102,582 commit pairs, constructs 2,308 self-verifying instances in 30 hours, and reduces environment size to 639GB while improving downstream pass@1.

\begin{table*}[htbp]
\centering

\footnotesize
\setlength{\tabcolsep}{6pt}
\begin{tabular}{lrrccr}
\toprule
\textbf{Dataset} & \textbf{\# Instances} & \textbf{\# Repos} & \textbf{Exec?} & \textbf{Source} & \textbf{Env. Size} \\
\midrule

R2E~\citep{jain2024r2e} & 0.25k & 137 & \cmark & Synth & 270\,GB \\
R2E-gym (Subset)~\citep{jain2025r2egymproceduralenvironmentshybrid} & 4.6k & 10 & \cmark & Synth & 4\,TB \\

SWE-bench-extra~\citep{badertdinov2024scaling} & 6.38k & 2k & \xmark & Real & -- \\
SWE-bench-train~\citep{jimenez2024swebench} & 19k & 37 & \xmark & Real & -- \\
SWE-fixer~\citep{xie2025swefixertrainingopensourcellms} & 115k & 856 & \xmark & Real & -- \\

SWE-Gym~\citep{pan2025trainingsoftwareengineeringagents} & 2.4k & 11 & \cmark & Real & 6\,TB \\
SWE-smith~\citep{yang2025swesmith} & 50k & 128 & \cmark & Synth & 295\,GB \\
SWE-rebench~\citep{badertdinov2025swerebenchautomatedpipelinetask} & 21.3k & 3.46k & \cmark & Real & -- \\
\midrule
\textbf{SWE-Next (ours)} & 2.3k & 3.97k & \cmark & Real & 639\,GB \\

\bottomrule
\end{tabular}
\caption{Dataset-scale comparison and environment size for open-source SWE training datasets. Most projects do not report their Env. Size. As a coarse lower-bound estimate, if a pipeline requires at least 100 distinct environments per repository, then scaling to thousands of repositories implies at least 10TB.
SWE-Next substantially reduces environment storage via reusable profiles, making large-scale task collection practical without requiring TB infrastructure. This improvement also greatly accelerates data collection: we processed all \textbf{3.97k repositories} and constructed the dataset in just \textbf{30 hours}.
}
\label{tab:env_size_comparison}
\end{table*}

\section{Related Work}
\label{sec:related_work}

\noindent\textbf{Repository-Level SWE Evaluation and Agents}
Repository-level software engineering has emerged as a central setting for evaluating LLM agents, largely driven by SWE-Bench and its Verified subset, which provide a standardized, test-verified harness for real-world issue resolution \citep{jimenez2024swebench,chowdhury2024swebenchverified}.
Unlike isolated code-generation benchmarks such as HumanEval \citep{chen2021evaluatinglargelanguagemodels} and MBPP \citep{austin2021programsynthesislargelanguage}, repository-level software engineering tasks require long-horizon reasoning over large codebases, dependency handling, and codebase exploration \citep{cai2026sweqaprorepresentativebenchmarkscalable}; issue-resolution settings further require execution in controlled repository environments.
These challenges have fueled rapid progress on repository-level agent systems and prompting workflows for GitHub issue resolution.
Broadly, existing SWE agents fall into two families: workflow-based systems decompose issue resolution into engineered stages such as localization, editing, and patch selection, which reduces horizon length but often requires substantial human priors and maintenance \citep{xia2024agentless,zhang2024autocoderoverautonomousprogramimprovement,antoniades2025swesearchenhancingsoftwareagents}; in contrast, general-purpose interactive agents rely more heavily on the underlying model to plan over longer horizons using tool calls and execution feedback \citep{yang2024sweagentagentcomputerinterfacesenable,wang2025openhandsopenplatformai}.
Across both families, a recurring limitation is the scarcity of high-signal training data and large-scale executable environments for open-weight repository-level agents, which helps explain the reliance on stronger proprietary models and prompting-heavy systems \citep{pan2025trainingsoftwareengineeringagents,jain2025r2egymproceduralenvironmentshybrid,yang2025swesmith}.

\noindent\textbf{Scalable and Verifiable Executable SWE Data Collection}
While SWE-Bench is central for evaluation, training data is harder to scale. Repository-specific environments are expensive and brittle to construct \citep{pan2025trainingsoftwareengineeringagents}, and reliable success signals typically require executing tests under controlled conditions, which is itself costly and sensitive to environment drift \citep{hu2025repo2run,badertdinov2025swerebenchautomatedpipelinetask}.
Recent work has therefore focused on building executable SWE data that is simultaneously realistic, verifiable, and scalable, spanning scalable environment construction and sandboxing \citep{hu2025repo2run}, automated task collection and decontaminated evaluation \citep{badertdinov2025swerebenchautomatedpipelinetask}, executable training environments with test-based signals \citep{wong2025confuciuscodeagentscalable}, procedural environment generation with verifiers and inference-time scaling \citep{jain2025r2egymproceduralenvironmentshybrid}, and large-scale task and trajectory synthesis for post-training \citep{yang2025swesmith}.
Across these systems, there is a recurring trade-off among scale, realism, and verification cost: synthetic perturbations can improve throughput but may weaken transfer to natural issues \citep{yang2025swesmith,hennicke2024mindgapsyntheticreal}, while human-authored issues, heavyweight per-task builds, or tightly curated evaluation pipelines improve validity but constrain scalability \citep{pan2025trainingsoftwareengineeringagents}.
SWE-Next is motivated by this trade-off and focuses on two complementary goals: improving supervision quality through execution-grounded filtering of real merged-PR candidates, and reducing collection overhead through reusable repo-quarter environment profiles.
Recent pipelines also use collected trajectories for post-training, but the quality of such rollouts remains fundamentally limited by the stability of the underlying environments and the reliability of instance construction \citep{xu2024benchmarkdatacontaminationlarge}.
This further motivates SWE-Next's emphasis on self-verifying task synthesis, strict submission gating, and low-overhead reusable environments as a practical foundation for scalable executable SWE data collection. Table~\ref{tab:env_size_comparison} summarizes the key differences between our method and prior work.

Recent concurrent efforts also report strong repository-level SWE performance by scaling task and trajectory corpora, broadening executable environment coverage, and pairing these resources with larger backbones or heavier post-training recipes\citep{tao2026swelegopushinglimitssupervised, chen2026sweuniversescalerealworldverifiable}.
These results are complementary to ours, but are not directly comparable because they operate under different model scales, data budgets, and training setups.
SWE-Next instead focuses on improving the collection pipeline itself, with the goal of improving supervision quality and collection efficiency through better filtering and reusable environment abstractions.

\section{Method}
\label{sec:method}

\subsection{Repository Ingestion and Execution-Grounded Task Synthesis}

We begin with a curated seed list of Python repositories on GitHub. We then filter the resulting repositories to prioritize active maintenance and executable test suites.
SWE-Next leverages the fact that real repositories already contain versioned fixes introduced by merged pull requests.
For each merged PR, we derive a candidate commit pair $(c_\text{base}, c_\text{merged})$.
Here, $c_\text{base}$ is the target-branch state immediately before the PR is merged, and $c_\text{merged}$ is the repository state after the PR is merged, regardless of whether the repository uses merge, squash, or rebase merging.

Given a synthesized environment, we execute a fixed repository test command on both commits and use execution-grounded filtering to determine whether the pair yields a valid training instance. We defer the detailed filtering rules and comparison logic to Section~\ref{sec:data_filtering}.
Retained pairs become self-verifying task instances mined from real pull requests: the ground-truth patch is the repository diff between $c_\text{base}$ and $c_\text{merged}$, and correctness is validated by execution rather than manual labeling, making the resulting instances compatible with SWE-Bench-style evaluation \citep{jimenez2024swebench}.

\subsection{Environment synthesis via reusable quarter profiles}
\begin{figure*}[t]
\centering
\includegraphics[width=0.9\textwidth]{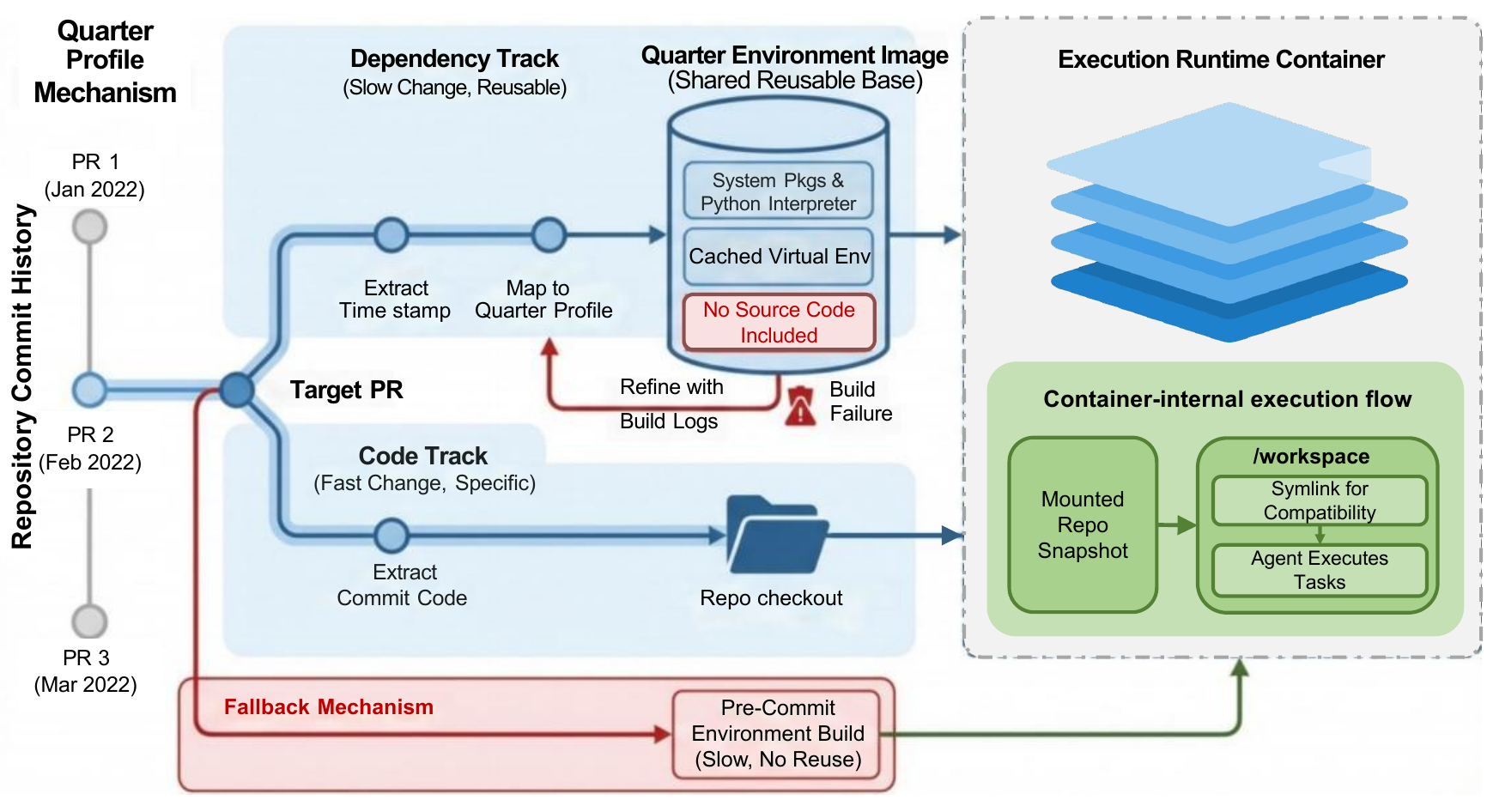}

\caption{Quarter profile mechanism for reusable environments.
We map each instance to a deterministic (repo, quarter) profile, build a shared quarter-env image that caches a virtual environment (without bundling repository source), and run per-commit repo checkouts in isolated workspaces while reusing the shared environment.
When the quarter-env image build fails, we fall back to per-commit environment builds.}
\label{fig:env_profile}
\end{figure*}

A major bottleneck in executable task synthesis is environment build cost and fragility: naively building a heavyweight Docker image for every commit pair repeats dependency installation and quickly becomes the dominant cost at scale \citep{hu2025repo2run}.
SWE-Next addresses this by separating execution into a reusable dependency layer and a commit-specific code snapshot.
Figure~\ref{fig:env_profile} provides a high-level overview of the quarter-profile mechanism and its fallback path.

\noindent\textbf{Quarter profiles.}
Within each repository, we deterministically map each commit timestamp to a coarse-grained \emph{quarter profile} of the form \texttt{repo\_name\_\{year\}Q\{quarter\}}, which approximates a dependency regime.
This design intentionally trades fine-grained per-commit specialization for reuse: in many real repositories, core dependencies evolve slowly relative to commit frequency, so a quarter-level partition captures most of the environment variation while enabling substantial amortization.
In our implementation, the profile identifier is passed through the runtime so both build-time and run-time logic can consistently select the same profile.

\noindent\textbf{Quarter environments (reusable images).}
For each profile, we build a shared quarter environment image that contains only the components that should be reused across commits, including system packages, a Python interpreter, and a cached virtual environment with pytest and profile-specific dependencies. These dependencies are inferred from repository metadata such as requirements files and project configuration, and can be further refined with an LLM when build logs reveal missing packages.
Crucially, quarter environments \emph{do not} bake repository source code into the image. Instead, the code is supplied at runtime, so the same quarter environment can be reused across many commits without rebuilding.

\noindent\textbf{Snapshot mounting, copy-on-start, and isolation.}
At execution time, SWE-Next mounts the commit-specific repository snapshot read-only, then copies it into a writable workspace before running tests or collecting trajectories.
This ``copy-on-start'' step ensures agent edits never modify the host snapshot while still allowing standard tooling like git diffs and test runs inside the container.
To preserve compatibility with existing repository test scripts, we link the shared quarter environment into the working directory, so repositories can reuse their original test command structure without modification.

\noindent\textbf{Robustness and fallback.}
Quarter profiling is an optimization rather than a hard assumption: when a quarter environment fails to build or is insufficient for a rare commit, SWE-Next falls back to per-commit environments for that commit to avoid dropping coverage.
When quarter-env builds fail, we optionally refine the profile specification using the Docker build error log, improving robustness over time while retaining the benefits of reuse.

SWE-Next can additionally build and tag per-commit final images for the subset of validated commits for reproducibility.
Appendix~\ref{app:env_profiles} provides implementation details and examples.

\subsection{Task packaging and synthetic issue statements}
After validation, SWE-Next converts each retained instance into a data row containing three components: the base commit, the ground-truth patch split into code and test diffs, and an executable environment specification.
To make instances usable for agent prompting without manual issue writing, we optionally generate a natural-language problem statement with an LLM from the available execution artifacts, including diff summaries and failing-test evidence. Prior work suggests that LLM-generated issue statements can be comparable to real issues for downstream training performance \citep{yang2025swesmith}.
Appendix~\ref{app:issue_generation} includes the problem-statement template, and Appendix~\ref{app:dataset_schema} documents the resulting dataset schema.

\subsection{Trajectory Generation and Baselines}
On top of the synthesized tasks, SWE-Next collects agent trajectories for post-training and analysis, following recent scalable SWE-agent pipelines \citep{jain2025r2egymproceduralenvironmentshybrid,yang2025swesmith}.
We use a rollout scaffold with two key defaults:
(i) Non-leaky prompting: by default, we do not inject curated \texttt{FAIL\_TO\_PASS}/\texttt{PASS\_TO\_PASS} lists into prompts, in order to avoid test leakage;
(ii) Strict submit gating: the agent is not allowed to submit unless it produces a non-empty code diff and runs at least one test command.
These choices are intended to improve the quality of collected supervision rather than simply increase trajectory volume.

To test whether this higher-signal collection strategy translates into better learning, we use the resulting trajectories for SFT and evaluate agent performance against recent baselines on SWE-Bench Verified~\citep{chowdhury2024swebenchverified} and SWE-Bench Lite~\citep{jimenez2024swebench}.

\section{Experiments}
\label{sec:experiments}

To assess the effectiveness of SWE-Next for training software engineering agents, we adopt an end-to-end fine-tuning protocol.
We (i) sample a subset of SWE-Next task instances, (ii) run an expert agent system to solve these tasks and collect execution-grounded trajectories, (iii) perform supervised fine-tuning (SFT) of a base/student model on the collected trajectories, and (iv) evaluate the resulting agent on standard SWE benchmarks.

\noindent\textbf{Agent System.}
Our trajectory-collection agent is based on the R2E-Gym agent system \citep{jain2025r2egymproceduralenvironmentshybrid}.
Concretely, we use the R2E-Gym scaffold, which builds on OpenHands-style tool-using agents and is further adapted for repository-level SWE tasks.
We inherit this scaffold and integrate it with SWE-Next instances and environments, enabling automated rollouts that produce step-by-step trajectories, including tool calls, observations, and edits.

\noindent\textbf{Models and Datasets.}
We seed SWE-Next with 3{,}971 Python repositories.
Across these repositories, SWE-Next executes 102{,}582 candidate instances and, after execution-based filtering, retains 2{,}308 final dataset instances. The repositories we selected have no overlap with the SWE-Bench dataset.
Given a subset of SWE-Next instances, we run the expert system to generate trajectories.
Unless otherwise noted, each rollout follows the same task interface, including repository state, task prompt, tool access, and the test-and-submit protocol, so that differences in downstream performance can be attributed to the collected data rather than to changes in the agent interface.
Using the expert agent system, we generate training trajectories with two expert models, Claude 4.5 Sonnet and GPT-5-mini, yielding 3.7k selected trajectories for SFT.
We retain two trajectory buckets for SFT: \emph{clean successes}, which end with a passing final test outcome, a real code edit, and consistently parseable passing earlier test evidence, and \emph{recovery successes}, which also end in a pass but include at least one earlier failing test before the final repair.
In both buckets, we bound trajectory length to control trace noise and keep collection costs stable.

For fairness in comparisons against prior work~\citep{jain2025r2egymproceduralenvironmentshybrid}, we also generate trajectories using claude-3-5-sonnet-20240620 under the same rollout scaffold and task interface.
We use the Qwen-2.5-Coder-Instruct 7B and 14B model families as base/student models.
We perform full-parameter supervised fine-tuning with LLaMA-Factory using a learning rate of $1.0\times10^{-5}$, a context length of 32{,}768 tokens, a global batch size of 8, and a maximum of 4 epochs; all SFT runs are conducted on 8 NVIDIA A100 GPUs.
When generating trajectories with expert models, we allow at most 40 interaction steps per rollout.
For student-model inference, we serve the fine-tuned models with vLLM, decode with temperature 0, and keep the same 40 step maximum so that collection and downstream evaluation use the same action budget.

\noindent\textbf{Evaluation Metrics and Benchmarks}
We evaluate downstream agent performance on SWE-Bench-Lite and SWE-Bench-Verified ~\citep{jimenez2024swebench}.
SWE-Bench-Lite is a reduced benchmark for faster iteration while preserving repository-level, test-verified evaluation.
SWE-Bench-Verified is a curated subset with higher-confidence instance validity, where solutions are judged by executing verification tests in controlled environments.
For each benchmark, we report pass@1, defined as the percentage of tasks for which the agent's final patch passes the benchmark evaluation.

\section{Execution-Grounded Data Filtering}
\label{sec:data_filtering}

We adopt a deliberately strict filtering policy because we treat executability, verifiability, and reproducibility as hard constraints, prioritizing data quality over recall.

\noindent\textbf{Repository-level filtering.}
To construct the repo list, we scrape the top 100 Python projects in each category from PyPI, extract their linked GitHub repositories, and remove duplicates.
We then filter the repositories to prioritize active maintenance and executable test suites. Appendix~\ref{app:repo_scope} summarizes the repository-level support conditions used in our pipeline.

\noindent\textbf{Commit-level pre-filtering.}
Before expensive execution, SWE-Next applies lightweight heuristics to avoid candidates with weak supervision and to control compute cost and noise.
We keep Python-source changes and drop documentation-only or obviously irrelevant edits.
We also reject commits with too many edited files or too many non-test changed lines.
Representative thresholds are reported in Appendix~\ref{app:commit_filtering}.
These pre-filters favor commit pairs that are small enough to localize, rich enough to support execution-grounded comparison, and cheap enough to evaluate at scale.

\noindent\textbf{Execution-grounded post-filtering.}
Given a synthesized environment, we execute a fixed repository test command on both commits and parse the resulting logs into structured per-test outcomes.
We compare only the overlapping test identifiers observed in both runs.
If no fully qualified per-test overlap exists, as can happen after refactors, we fall back to file-level matching by grouping tests under their test files.
We further classify each candidate into execution types, including setup failure, test-run failure, \textsc{NewCommitNotBetter}, and \textsc{NewCommitBetter}.
A pair is labeled as \textsc{NewCommitBetter} if the merged commit turns at least one previously non-passing test into \textsc{Passed} and introduces no regressions under the same test command.
Only \textsc{NewCommitBetter} candidates proceed to dataset construction. Appendix~\ref{app:exec_types} formalizes the execution-type taxonomy and comparison rules.

In the run described in this paper, SWE-Next starts from a seed set of 3{,}971 repositories, whose domain composition is shown in Appendix Figure~\ref{fig:repo_domain_pie}, executes 102{,}582 candidate commit pairs, and retains 2{,}308 self-verifying instances for a yield of \SwenextFinalYieldPct{}\%.
The retained instances span 311 repositories. Appendix~\ref{app:repo_list} provides the full repository list together with one-sentence summaries in Table~\ref{tab:newcommitbetter_repos}.
Although the yield is low, this is expected in real repositories: 74.5\% of candidates do not improve test behavior, 2.5\% fail during setup, and 20.8\% fail during test execution.

\noindent\textbf{Why many repositories yield zero instances.}
The discarded candidates span several distinct failure modes even within repositories that otherwise look suitable for mining.
For example, the black repository yields zero \textsc{NewCommitBetter} instances overall: across the mined candidates, it contains 293 \textsc{NewCommitNotBetter} cases and 1 setup failure, but no execution-grounded strict improvement.
A representative candidate is the change \emph{``Preserve line endings when formatting a file in place''}, which is labeled \textsc{NewCommitNotBetter} because both commits execute one comparable test, yet none of the overlapping tests improve from non-passing to passing.
As a result, many repositories yield zero instances not because they are unimportant, but because no candidate in that repository survives all three layers of filtering. Additional system statistics are reported in Appendix~\ref{sec:apx_system_stats}.

\begin{table*}[htbp]
\centering
\small
\setlength{\tabcolsep}{6pt}
\begin{tabular}{llcccc}
\toprule
\textbf{Pipeline} & \textbf{Backbone} & \textbf{System} & \textbf{\# Train Trajs} & \textbf{Verified} & \textbf{Lite} \\
\midrule
\multicolumn{6}{l}{\textit{Comparable trajectory-SFT baselines}} \\
Qwen2.5-Coder-Instruct   & 7B  & OpenHands & -- & 1.8 & 1.0 \\
R2E-Gym   & 7B  & OpenHands & 3.3k & 14.4 & 11.3 \\
SWE-smith & 7B  & SWE-agent & 2k   & 15.2 & 11.7 \\
\textbf{SWE-Next (ours)} & \textbf{7B}  & OpenHands & 3.7k & \textbf{17.4} & \textbf{13.7} \\
Qwen2.5-Coder-Instruct   & 14B  & OpenHands & -- & 4.0 & 2.7 \\
R2E-Gym   & 14B & OpenHands & 3.3k & 26.8 & 20.7 \\
\textbf{SWE-Next (ours)} & \textbf{14B} & OpenHands & 3.7k & \textbf{30.0} & \textbf{23.3} \\
\midrule
\multicolumn{6}{l}{\textit{Larger-scale or differently configured references}} \\
SWE-Gym   & 32B        & OpenHands & 491  & 20.6 & 15.3 \\
SWE-fixer & 72B        & SWE-Fixer & 110k & 32.8 & 24.7 \\
Qwen2.5-Coder Instruct & 32B        & OpenHands & --   & 7.0   & 3.0 \\
Qwen3-A22B             & 235B-A22B  & OpenHands & --   & 34.4   & -- \\
\bottomrule
\end{tabular}
\caption{Downstream agent pass@1 resolve rates on SWE-Bench-Verified and SWE-Bench-Lite, with the number of training trajectories used for SFT.
We do not compare against systems that use verifiers or multiple attempts at test time.\label{tab:main_results}}
\end{table*}

\section{Main Results}
\label{sec:results}

Table~\ref{tab:main_results} summarizes pass@1 on SWE-Bench-Lite and SWE-Bench-Verified.
Under matched student sizes on both benchmarks, SWE-Next consistently improves downstream agent performance over prior trajectory-collection and task-synthesis pipelines.
The improvements are more pronounced on SWE-Bench-Lite, indicating stronger transfer in faster-iteration settings.
Notably, SWE-Next (14B) also exceeds SWE-Gym (32B) on SWE-Bench-Verified, suggesting that higher-signal data construction and trajectory defaults can rival model scaling in SWE-Bench evaluation.

\noindent\textbf{Data scale vs.\ downstream performance.}
Prior work suggests that collecting more agent trajectories for SFT generally leads to stronger downstream SWE-agent performance \citep{jain2025r2egymproceduralenvironmentshybrid,pan2025trainingsoftwareengineeringagents}.
SWE-Next, however, focuses not only on scaling trajectory collection, but also on improving trajectory quality through execution-grounded filtering and high-signal rollout defaults.
In the 7B/14B student regime, SWE-Next achieves higher pass@1 with fewer or comparable numbers of training trajectories, suggesting that data quality matters as much as, if not more than, raw trajectory volume.

\noindent\textbf{Efficient Environments.}
\label{sec:system_advantages}To contextualize executable environment size, Table~\ref{tab:env_size_comparison} compares environment sizes across representative open-source SWE training datasets.
For SWE-Next, we report an environment size of 639GB for the run described in this paper.
Many projects do not report their environment size. As a coarse lower-bound estimate of scaling without reuse, even a modest assumption of 100 distinct environments per repository would already imply at least 10TB of environment images when scaled to thousands of repositories.
As a further systems estimate, if we disable repo-quarter profiles and instead materialize one environment per executed candidate, the full 102{,}582-candidate collection run would require approximately 30.8\,TB of environment images under a conservative 300\,MB-per-image assumption.
While this is only an estimate, it highlights why environment reuse is essential for scalable executable SWE data collection.
Our repo-quarter profile mechanism therefore reduces storage and build overhead while accelerating large-scale data collection, lowering the resource barrier for smaller academic labs and industry teams to construct executable SWE training datasets.

\noindent\textbf{High-signal trajectories via strict interfaces and non-leaky prompts.}
We also provide a rollout interface with submission gating that requires agents to produce a valid code diff before submitting, along with prompts that encourage them to reproduce the bug, run tests, and verify the fix before submission.
In our collection run, \SwenextTrajSuccessPct{}\% of trajectories succeed. Among the successful trajectories, agents execute at least one test command in \SwenextTrajRanTestPct{}\% of cases and run a reproduction script in \SwenextTrajRanReproPct{}\%.
Importantly, we do not expose ground-truth test lists in the agent prompt, preventing label leakage while still enabling verifiable evaluation.

\begin{table}[htbp]
\centering
\caption{Performance gains are not driven by a stronger trajectory generator, but by SWE-Next's execution-grounded task synthesis and supervision.}
\small
\setlength{\tabcolsep}{6pt}
\begin{tabular}{lcc}
\toprule
\textbf{SFT data} & \textbf{SWE-Bench Verified} & \textbf{SWE-Bench Lite} \\
\midrule
R2E-Gym & 7.8\% & 5\% \\
SWE-Next (ours) & \textbf{8.2\%} & \textbf{6.67\%} \\
\bottomrule
\end{tabular}

\label{tab:traj_generator_ablation}
\end{table}

\noindent\textbf{Ablation: Controlling for the trajectory generator.}
Our training trajectories are generated with GPT-5-mini and Claude 4.5 sonnet, while the released R2E-Gym trajectories are generated with Claude 3.5 Sonnet.
To ensure a fair comparison and isolate the effect of the framework from the different trajectory generator, we perform a controlled study on 800 randomly sampled SWE-Next task instances.
For each framework we select 400 successful trajectory and matched trajectory sets using the same backbone model (Claude 3.5 Sonnet).
We then SFT the same 7B model on each 400 successful trajectory set under identical training hyperparameters, and evaluate on SWE-Bench Verified and SWE-Bench Lite.
As shown in Table~\ref{tab:traj_generator_ablation}, the model trained on SWE-Next trajectories consistently outperforms the model trained on R2E-Gym trajectories on both benchmarks, suggesting that our gains do not stem from using a stronger trajectory generator but from SWE-Next's execution-grounded task synthesis and supervision.

In conclusion, by combining reusable quarter environments with high-signal, execution-grounded trajectories, SWE-Next makes large-scale data collection both faster and higher quality, enabling efficient synthesis without sacrificing downstream performance.

\section{Failure Analysis}
\label{sec:failure_analysis}
We analyze failed trajectories in our collection run and classify them using signals already emitted by the pipeline. This yields four practically meaningful categories:
(i) residual env/harness failure, where the trajectory is blocked by unresolved pytest or environment issues;
(ii) submit-gating or no-valid-diff failure, where the run ends without a valid final repository diff;
(iii) search/localization failure, where the agent modifies a seemingly relevant location, but final verification still misses part of the target test set, suggesting that it captures only part of the bug rather than the full issue; and
(iv) patch quality failure, where the expected tests are executed but at least one target test still fails.
Table~\ref{tab:failure_type_decomp} reports the resulting distribution.

\begin{table}[htbp]
\centering

\small
\setlength{\tabcolsep}{4pt}
\begin{tabular}{lr}
\toprule
\textbf{Failure Type} & \textbf{\% of failures} \\
\midrule
Residual env/harness        & 2.4\% \\
Search/localization         & 8.7\% \\
Patch quality               & 75.7\% \\
Submit gating / no valid diff & 13.2\% \\
\bottomrule
\end{tabular}
\caption{Failure-type decomposition over failed SWE-Next trajectories. With reusable repo-quarter profiles and runner fixups, residual env or harness failures are rare.}
\label{tab:failure_type_decomp}
\end{table}

\noindent\textbf{Most failed runs are still substantive debugging attempts.}
Thanks to our reusable repo-quarter profile mechanism, residual env or harness failures are rare, accounting for only 2.4\% of all failures.
Our strict submission gating also keeps failed trajectories high-signal rather than speculative: among failed runs, 86.0\% execute at least one test command, 67.2\% run a reproduction script, 86.8\% edit repository files, and 78.4\% reach finish.
Moreover, 87.4\% of failures end by hitting the step budget rather than crashing at runtime, indicating that the dominant bottleneck is not infrastructure, but agent debugging and patch quality under a strict execution-grounded interface.
The largest failure category is patch quality failure, accounting for 75.7\% of all failures.
These are not random breakdowns. The verifier can parse the full expected test set, but one or more target tests still fail.
In most such cases, the agent appears to identify the correct subsystem and produce a meaningful patch, yet still misses a state invariant, an edge case, or a semantic detail required to match the gold behavior.
Search and localization failures are much less frequent at 8.7\%. These failures typically occur when the agent identifies only part of the relevant code or addresses only part of the bug, leaving the full target behavior unsatisfied.
Submit-gating failures account for 13.2\% of failures and are also informative, because the pipeline explicitly records runs without a valid final diff instead of silently accepting speculative answers.
Representative case studies for all four failure buckets are provided in Appendix~\ref{app:failure_case_studies}.

\section{Discussion}
\noindent\textbf{Limitations.}
\label{sec:limitations}First, SWE-Next currently focuses on Python repositories. While its core ideas are not specific to Python, extending the pipeline to other languages remains future work.
Second, we do not yet explore trajectory generation at the largest possible scale. Due to budget and compute constraints, our collection run is designed primarily to validate the SWE-Next framework rather than to maximize raw trajectory count. Our current results already show gains in efficiency and downstream performance, but larger-scale trajectory collection remains to be explored.

\noindent\textbf{Conclusion.}
\label{sec:conclusion}In this work, we introduced SWE-Next, a framework for scalable executable software engineering data collection that improves both supervision quality and systems efficiency. By mining real merged pull requests and retaining only commit pairs that produce strict execution-grounded improvements without regressions, SWE-Next turns repository history into self-verifying tasks and high-signal training trajectories, thereby decoupling supervision quality from raw data volume. By combining strict execution-based validation with reusable repo-quarter environment profiles, SWE-Next also makes large-scale executable data collection far more practical and accessible for research. Our findings suggest that progress in SWE agents may depend not only on larger training corpora, but on building data pipelines that simultaneously improve supervision quality, reduce systems overhead, and preserve reproducibility. We believe this direction offers a path toward more efficient, reproducible, and trustworthy post-training for software engineering agents.

\newpage
\bibliography{colm2026_conference}
\bibliographystyle{colm2026_conference}

\appendix

\section{Additional Method Details}
\label{app:method_details}

\subsection{Repository Scope, Selection Rules, and Dataset Scale}
\label{app:repo_scope}
\textbf{What counts as a ``supported repository''.}
SWE-Next currently targets a curated set of Python repositories for which we provide end-to-end support (repo ingestion, environment setup, log parsing, and a stable test command).
The list is intentionally curated to keep the pipeline reproducible and to avoid silent failures caused by unusual build systems or non-standard test harnesses.

\begin{swenextbox}{Supported-Repository Selection Rules}
\small
We include repositories only when the full pipeline can run end to end with a stable execution contract:
\begin{itemize}
  \item \textbf{Stable Python test runner:} typically \texttt{pytest} invokable from the repository root.
  \item \textbf{Usable bug-fix history:} merged commits or PRs with enough test signal to support execution-grounded comparison.
  \item \textbf{Containerizable setup:} system dependencies and Python dependencies can be expressed in a reproducible image recipe.
  \item \textbf{Manageable runtime:} setup and test execution fit within CI-style time and resource budgets.
\end{itemize}
\end{swenextbox}

\textbf{Additional System Statistics}
\label{sec:apx_system_stats}

\begin{table}[htbp]
\centering

\footnotesize
\setlength{\tabcolsep}{5pt}
\begin{tabular}{lp{0.33\columnwidth}}
\toprule
\textbf{Metric} & \textbf{Value} \\
\midrule
Candidates executed & \SwenextCandidatesExecuted{} \\
Final instances & \SwenextFinalInstances{} (\SwenextFinalYieldPct{}\% yield) \\
Candidates with unchanged test behavior & 74.5\% \\
Candidates failing during setup & 2.5\% \\
Candidates failing during test execution & 20.8\% \\
Unique repos in final & \SwenextFinalRepos{} \\
Repo-quarter profiles & \SwenextRepoQuarterProfiles{} \\
Unique env signatures & \SwenextUniqueEnvSignatures{} \\
Quarters covered (date span) & 55 (\SwenextCommitDateMin{}--\SwenextCommitDateMax{}) \\
Total trajectories & \SwenextTrajTotal{} \\
Selected trajectories & \SwenextTrajSuccess{} \\
Successful: ran $\geq 1$ test command & \SwenextTrajRanTestPct{}\% \\
Successful: ran a reproduction script & \SwenextTrajRanReproPct{}\% \\

\bottomrule
\end{tabular}
\caption{Additional system statistics for the SWE-Next run described in this paper.}
\label{tab:swenext_system_stats_apx}
\end{table}

\subsection{Commit Filtering Heuristics}
\label{app:commit_filtering}
SWE-Next uses lightweight heuristics to (i) avoid commits with weak supervision (no tests) and (ii) control compute cost and noise.
We filter candidates before expensive execution whenever possible, and we apply additional constraints after execution.

\begin{swenextbox}{Conceptual Candidate-Filtering Protocol}
\small
SWE-Next uses lightweight pre-filters to avoid spending execution budget on commit pairs with weak supervision or excessive noise.
\begin{itemize}
  \item \textbf{Patch-size caps:} reject commits with too many edited files or too many non-test changed lines.
  \item \textbf{Language and file filters:} keep Python-source changes and drop documentation-only or obviously irrelevant edits.
  \item \textbf{Execution-worthiness filters:} prioritize commits with usable test evidence before running the expensive verifier.
\end{itemize}
Representative knobs include \texttt{max\_num\_non\_test\_files = 5}, \texttt{max\_num\_non\_test\_edited\_lines = 200}, \texttt{max\_patch\_length = 10000}, \texttt{keep\_only\_python\_commits = true}, and \texttt{keep\_only\_small\_commits = true}.
\end{swenextbox}

\begin{swenextbox}{Execution-Grounded Post-Filtering Protocol}
\small
After execution, SWE-Next keeps only commit pairs that produce a stable, comparable verification signal:
\begin{itemize}
  \item \textbf{Drop setup failures:} remove candidates whose environment cannot be built or initialized reliably.
  \item \textbf{Drop invalid test runs:} remove candidates whose test invocation crashes, times out, or cannot be parsed into comparable outcomes.
  \item \textbf{Retain strict improvements only:} keep a candidate only when at least one test moves from fail to pass and no previously passing test regresses.
\end{itemize}
This stage converts raw execution outcomes into the labels described in Appendix~\ref{app:exec_types}; only \textsc{NewCommitBetter} instances are released in the final dataset.
\end{swenextbox}

\subsection{Repository Installation and Runtime Bring-up}
\label{app:repo_installation}
SWE-Next standardizes environment bring-up inside Docker so that verification and agent runs are reproducible.
At a high level, each run requires:
(i) a base container image with system dependencies and a Python interpreter,
(ii) a repo snapshot mounted into the container, and
(iii) a deterministic setup procedure to install dependencies.

\subsection{Test Extraction / ``Test Generation''}
\label{app:test_generation}
SWE-Next relies on execution-grounded tests rather than purely LLM-generated unit tests.
For each candidate, we aim to execute a \emph{minimal but informative} test suite to determine whether the new commit improves behavior.

\textbf{Test-suite construction.}
We extract a small set of tests from repository history and/or from the commit's test diff and store them as artifacts alongside the instance.
At runtime, SWE-Next materializes these tests inside the container (e.g., under a dedicated folder such as \texttt{r2e\_tests/}) and invokes a standard test command.

\subsection{Problem Statement Generation Template}
\label{app:issue_generation}
To support agent prompting without manual issue writing, SWE-Next can generate a short natural-language problem statement using an LLM conditioned on execution artifacts.
This is optional and can be disabled for fully non-LLM dataset construction.

\begin{swenextbox}{Problem-Statement Generation Prompt (High-Level)}
\small
The issue generator receives execution artifacts for a commit pair and is instructed to produce a concise GitHub-issue-style statement with three elements:
\begin{itemize}
  \item \textbf{Observed failure:} what breaks on the base commit.
  \item \textbf{Expected behavior:} what should happen after the fix.
  \item \textbf{Relevant constraints:} only hints that are directly supported by the diff or logs.
\end{itemize}
The output is constrained to a single issue block of the form \texttt{[ISSUE] ... [/ISSUE]} so that the generated statement can be extracted deterministically.
\end{swenextbox}

\subsection{Trajectory Collection Interface and Submission Gating}
\label{app:trajectory_protocol}
SWE-Next's second contribution is not only task synthesis, but also a high-signal rollout interface that makes collected traces more faithful to downstream SWE-Bench-style evaluation.

\begin{swenextbox}{High-Signal Rollout Protocol (High-Level)}
\small
Each trajectory is collected under a fixed interface designed to encourage evidence-driven debugging rather than speculative patching:
\begin{itemize}
  \item \textbf{Non-leaky task prompt:} the agent receives the repository state and task statement, but not the curated \texttt{FAIL\_TO\_PASS}/\texttt{PASS\_TO\_PASS} lists.
  \item \textbf{Mandatory evidence gathering:} the prompt and tool scaffold encourage the agent to inspect files, reproduce the bug, and run tests before final submission.
  \item \textbf{Strict submit gating:} a submission is accepted only if the agent produces a non-empty code diff; empty or purely speculative finishes are recorded as failures.
  \item \textbf{Verifiable logging:} tool calls, observations, diffs, and execution outcomes are all retained as structured trajectory artifacts for later filtering and post-training.
\end{itemize}
\end{swenextbox}

\subsection{Patch Minimization and Patch Splitting}
\label{app:patch_minimization}
SWE-Next keeps patches training- and evaluation-friendly by (i) filtering out overly large candidates and (ii) splitting patches into \emph{code} and \emph{test} components, following the SWE-Bench convention.

\textbf{Patch splitting.}
Given a commit diff, we partition file diffs into:(1)non-test patch: source-code changes used as the ground-truth fix;(2)test patch: test-only changes (stored separately).

\subsection{Reusable Environment Profiles and Quarter Environments}
\label{app:env_profiles}
This appendix explains how the ``reusable environment profiles'' mechanism is implemented.

\textbf{Motivation.}
Building a final Docker image per commit is expensive and can accumulate large local build caches.
SWE-Next therefore amortizes environment building by reusing a shared \emph{quarter environment} across many commits.

\begin{swenextbox}{Repo-Quarter Profile Construction}
\small
SWE-Next derives a reusable environment profile from the repository identifier and commit timestamp:
\begin{itemize}
  \item \textbf{Profile key:} \texttt{profile = <repo\_name>\_<year>Q<quarter>}
  \item \textbf{Example:} \texttt{black + 2022-05-01} $\rightarrow$ \texttt{black\_2022Q2}
  \item \textbf{Design rationale:} the quarter granularity captures slow-moving dependency regimes while avoiding one expensive environment build per commit.
\end{itemize}
\end{swenextbox}

\textbf{Quarter environment build.}
For each profile, we build a Docker image that contains:
(i) system packages (apt deps),
(ii) a Python interpreter,
(iii) cached/pinned Python dependencies needed by that repo profile.
The profile Dockerfile is stored as an artifact so the environment can be rebuilt.

\textbf{Repo snapshot mounting and copy-on-start.}
At runtime, the host-side repo snapshot for a specific commit is mounted into the container read-only.
The container then copies the snapshot to a writable workspace before agent edits and test runs. This prevents accidental writes to the host filesystem and keeps runs isolated.

\subsection{Commit Execution Types: Definitions}
\label{app:exec_types}
SWE-Next classifies each candidate commit pair using execution-grounded labels.
Let $\mathcal{R}(c)$ denote the parsed outcomes of executing the test command at commit $c$.
We parse outputs into per-test statuses (e.g., \texttt{PASSED}, \texttt{FAILED}), keyed by a stable identifier such as \texttt{file::class::test}.

\textbf{Setup/test-run failures.}
\begin{itemize}
  \item \textbf{Setup failure}: environment creation fails (dependency install, missing binaries, etc.).
  \item \textbf{Test-run failure}: tests run but logs cannot be parsed or no comparable tests are found.
\end{itemize}

\textbf{\textsc{NewCommitNotBetter}.}
The new commit is \emph{not better} when:
\begin{itemize}
  \item no test transitions from failing to passing between $c_\text{base}$ and $c_\text{merged}$, and
  \item there is no clear improvement under the comparison rule.
\end{itemize}

\textbf{\textsc{NewCommitBetter}.}
The new commit is \emph{better} if at least one comparable test improves and none regress:
\begin{swenextbox}{Execution-Grounded Commit Comparison Rule}
\small
For two comparable test outcomes, SWE-Next defines
\[
\mathrm{Improved} = \{ t \in \mathrm{CommonTests} \mid \mathcal{R}(c_{\text{base}})[t] \neq \texttt{PASSED} \ \land\  \mathcal{R}(c_{\text{merged}})[t] = \texttt{PASSED} \},
\]
\[
\mathrm{Regressed} = \{ t \in \mathrm{CommonTests} \mid \mathcal{R}(c_{\text{base}})[t] = \texttt{PASSED} \ \land\  \mathcal{R}(c_{\text{merged}})[t] \neq \texttt{PASSED} \}.
\]
We label a candidate as \textsc{NewCommitBetter} iff $|\mathrm{Improved}| > 0$ and $|\mathrm{Regressed}| = 0$.
\end{swenextbox}
When there is no overlap in fully-qualified test identifiers, we fall back to file-level comparison (grouping tests by test file) to handle refactors.

\subsection{Dataset Schema and Example Instance}
\label{app:dataset_schema}
SWE-Next emits SWE-Bench-like JSONL instances plus per-instance sidecar artifacts for reproducibility (dockerfile, execution logs, parsed diffs).

\begin{swenextbox}{Released Instance Format (Core Example)}
\small
Each released JSONL row is SWE-Bench-like and includes a compact task definition together with the artifacts needed for deterministic replay:
\begin{itemize}
  \item \textbf{Task identity:} \texttt{repo}, \texttt{instance\_id}, \texttt{base\_commit}, \texttt{commit\_hash}
  \item \textbf{Reference patches:} \texttt{patch}, \texttt{test\_patch}
  \item \textbf{Task statement:} \texttt{problem\_statement}
  \item \textbf{Environment artifact:} \texttt{dockerfile} and the resolved \texttt{docker\_image}
\end{itemize}
\end{swenextbox}

\textbf{Dataset-card summary.}
Table~\ref{tab:dataset_card_fields} summarizes the top-level fields in the released SWE-Next JSONL file. Large structured artifacts are serialized as JSON strings so that each row remains self-contained.
These fields are preserved for reproducibility and offline verification; during rollout, the agent prompt only receives \texttt{problem\_statement}, not the structured commit, execution, or expected-output payloads.

\begin{longtable}{p{0.28\linewidth}p{0.68\linewidth}}
\toprule
\textbf{Field} & \textbf{Meaning} \\
\midrule
\endfirsthead
\toprule
\textbf{Field} & \textbf{Meaning} \\
\midrule
\endhead
\midrule \multicolumn{2}{r}{\small Continued on next page} \\
\endfoot
\bottomrule
\noalign{\vskip 7pt}
\caption{Top-level fields in the released SWE-Next JSONL dataset.\label{tab:dataset_card_fields}}
\endlastfoot
\multicolumn{2}{l}{\textit{Identity and provenance}} \\
repo & Canonical repository name in \texttt{owner/name} format. \\
repo\_name & Normalized repository identifier used by the pipeline for local naming and artifact organization. \\
repo\_key & Filesystem-safe repository key used in image and artifact naming. \\
instance\_id & Unique task identifier for the commit pair. \\
base\_commit & Pre-fix commit SHA used as the buggy starting point. \\
commit\_hash & Post-fix commit SHA whose behavior defines the target solution. \\
created\_at & Timestamp of the post-fix commit, carried over as task provenance metadata. \\
\midrule
\multicolumn{2}{l}{\textit{Task definition and executable labels}} \\
patch & Non-test unified diff between \texttt{base\_commit} and \texttt{commit\_hash}; this is the reference code patch. \\
test\_patch & Test-only unified diff for the same commit pair. \\
problem\_statement & Natural-language task statement synthesized from the commit diff and execution evidence. \\
hints\_text & Optional hint field kept for SWE-Bench compatibility. \\
FAIL\_TO\_PASS & List of target tests that fail on \texttt{base\_commit} and pass on \texttt{commit\_hash}. \\
PASS\_TO\_PASS & List of regression-guard tests that pass on both commits. \\
exec\_type & Execution-grounded label assigned during filtering; for this release all rows are \texttt{NEW\_COMMIT\_BETTER}. \\
\midrule
\multicolumn{2}{l}{\textit{Environment and verification artifacts}} \\
environment\_setup\_commit & Commit used to synthesize the runtime environment. \\
docker\_image & Resolved container image tag for executing and verifying the instance. \\
dockerfile & Self-contained Docker build recipe used to produce \texttt{docker\_image}. \\
version & Optional version tag for the repository state. \\
difficulty & Optional difficulty annotation placeholder. \\
\midrule
\multicolumn{2}{l}{\textit{Generation metadata}} \\
prompt & Input prompt used to synthesize \texttt{problem\_statement} from commit and execution metadata; it is not the downstream agent prompt. \\
parsed\_commit\_content & JSON-encoded structured commit object containing file diffs, commit message/date, and commit metadata. \\
execution\_result\_content & JSON-encoded execution record containing setup/test return codes, stdout/stderr, and embedded test-file reconstruction artifacts. \\
expected\_output\_json & JSON-encoded mapping from test identifier to expected status on the post-fix commit. \\
\end{longtable}

\section{Failure Bucket Case Studies}
\label{app:failure_case_studies}

\noindent\textbf{Residual env/harness failure.}
On the pex-tool/pex instance \emph{``Importing dependencies from a non-executable venv PEX breaks standard library imports''}, the agent inspects and edits pex.py, and reruns run\_tests.sh.
The run reaches final verification, but the remaining target test test\_standard\_library\_is\_included still fails with ModuleNotFoundError: No module named 'contextvars'.
This is representative of the residual env/harness bucket: the trajectory makes a code change, but unresolved interpreter or environment issues still block clean verification.

\noindent\textbf{Submit-gating / no-valid-diff failure.}
On the 23andMe/Yamale instance \emph{``String constraints do not correctly validate equality, prefix, suffix, regex, and case-insensitive exclusions''}, the agent inspects validators.py and constraints.py, runs pytest -q, and attempts to inspect run\_tests.sh.
It reaches the finish step with a textual summary of the issue, but produces no final repository diff.
The verifier can still parse all 19 expected tests, yet only 11 behave as required, so the run is recorded as a submit-gating failure rather than accepted as a speculative attempt.

\noindent\textbf{Search/localization failure.}
On the Blazemeter/taurus instance \emph{``PassFailStatus does not trigger shutdown on cumulative data criteria during shutdown phase''}, the agent localizes to passfail.py, modifies the logic in aggregated\_second, writes and executes reproduce\_issue.py, reruns test\_passFailStatus.py, and finally calls run\_tests.sh.
The patch is directionally sensible: the verifier parses 5 passing expected tests.
However, 4 expected tests remain unmatched in the final evidence.
This is representative of the search/localization bucket: the agent fixes one observed path, but its verification remains incomplete relative to the full behavioral surface encoded by the expected test set.

\noindent\textbf{Patch quality failure.}
On the Blazemeter/taurus instance \emph{``KPISet fails to handle response times length limit causing attribute errors and incorrect merging behavior''}, the agent localizes to aggregator.py, writes and executes reproduce\_issue.py, reruns run\_tests.sh, and submits a non-empty patch.
Unlike the search/localization case, the verifier parses the full expected test set: all 6 target tests are present in the final evidence.
Nevertheless, one key test, TestConsolidatingAggregator::test\_set\_rtimes\_len, still fails, leaving the run at 5/6 expected passes.
This is representative of the patch-quality bucket: the agent reaches the right subsystem and produces a plausible edit, but the final semantics are still not fully correct.

\section{Seed Repository Domain Composition}
\label{app:repo_domain}

\begin{figure}[htbp]
\centering
\includegraphics[width=0.5\linewidth]{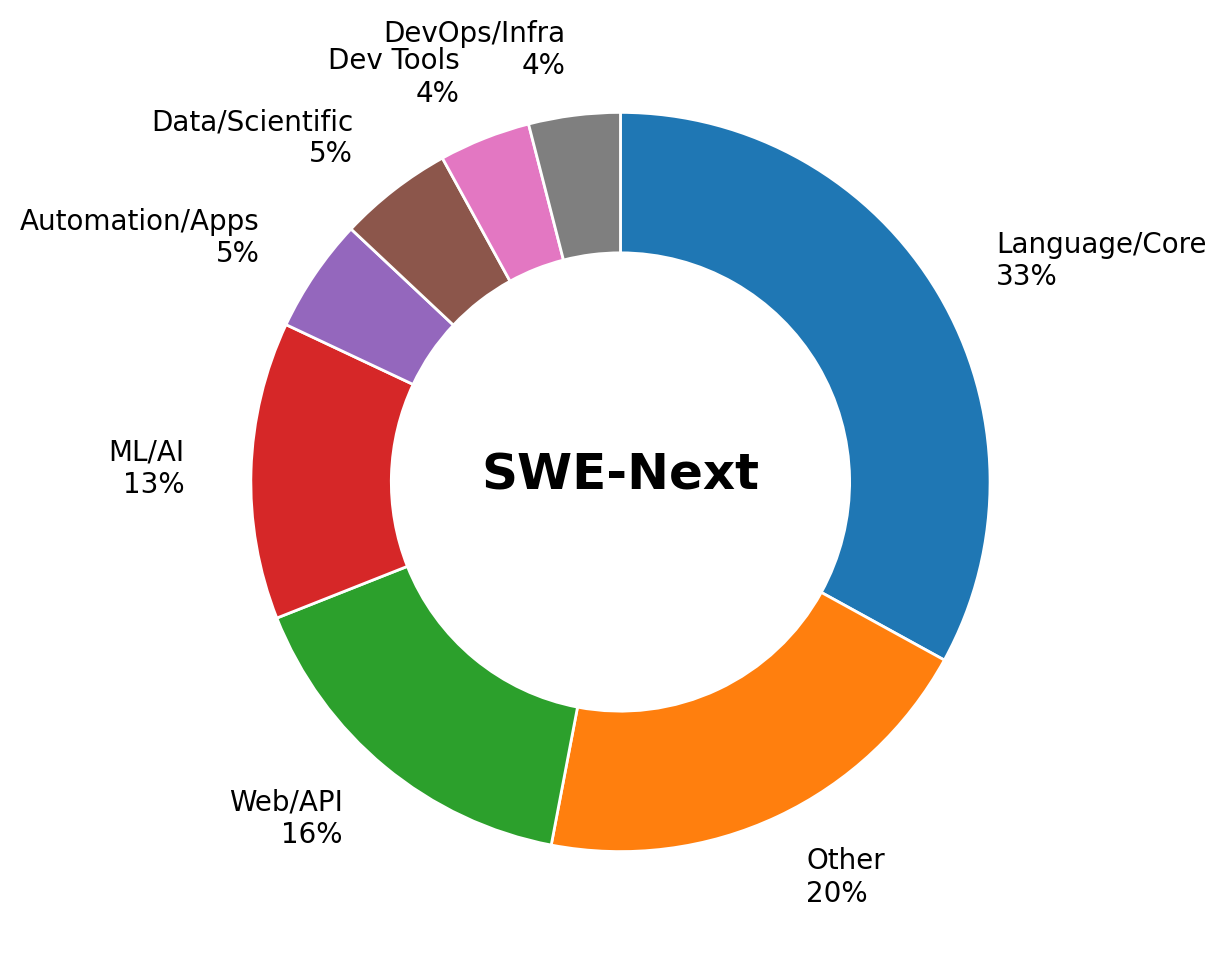}
\caption{Domain composition of the 3{,}971 seeded repositories, inferred from GitHub repository topics and descriptions. The corpus spans diverse software domains.}
\label{fig:repo_domain_pie}
\end{figure}

\section{Repository List (NewCommitBetter)}
\label{app:repo_list}
Table~\ref{tab:newcommitbetter_repos} lists the 311 repositories that contribute at least one \textsc{NewCommitBetter} instance in our final dataset, grouped by coarse domain with one-sentence descriptions derived from GitHub metadata.
\input{tables/newcommitbetter_repo_descriptions.tex}

\section{Author Contributions and Acknowledgments}

All authors participated in discussions throughout the project and contributed ideas that shaped the direction of the work. Jiarong Liang implemented the full SWE-Next framework, carried out dataset, trajectory collection and supervised fine-tuning, and conducted the majority of the evaluation. Zhiheng Lyu proposed the initial idea for the project and contributed through continued discussion and insightful feedback during its development. Zijie Liu contributed to part of the evaluation and to the writing of the paper. Xiangchao Chen explored trajectory generation using MinMax-based models. Ping Nie and Kai Zou provided computational resources and API support. Wenhu Chen helped refine the research problem and provided important insights into the project design, which substantially improved the effectiveness of the subsequent analysis.

We also thank Yuansheng Ni and Yubo Wang for valuable experience, suggestions, and feedback on both the technical challenges of the project and the writing of the paper.

\end{document}

%% file: tables/newcommitbetter_repo_descriptions.tex
\begin{longtable}{p{0.30\linewidth}p{0.66\linewidth}}
\toprule
\textbf{Repository} & \textbf{Description} \\
\midrule
\endfirsthead
\toprule
\textbf{Repository} & \textbf{Description} \\
\midrule
\endhead
\midrule \multicolumn{2}{r}{\small Continued on next page} \\
\endfoot
\bottomrule
\noalign{\vskip 7pt}
\caption{Repositories that contribute \textsc{NewCommitBetter} instances in our dataset, grouped by coarse domain. Descriptions are derived from GitHub repository metadata.\label{tab:newcommitbetter_repos}}
\endlastfoot
\multicolumn{2}{c}{\textit{Dev Tools}} \\
\midrule
avocado-framework/avocado & Avocado is a set of tools and libraries to help with automated testing. \\
Blazemeter/taurus & Automation-friendly framework for Continuous Testing by. \\
etianen/logot & Test whether your code is logging correctly . \\
pylint-dev/astroid & A common base representation of python source code for pylint and other projects. \\
pytest-dev/pyfakefs & Provides a fake file system that mocks the Python file system modules. \\
pytest-dev/pytest & The pytest framework makes it easy to write small tests, yet scales to support complex functional testing. \\
snok/flake8-type-checking & Flake8 plugin for managing type-checking imports \& forward references. \\
sondrelg/flake8-typing-only-imports & Flake8 plugin for managing type-checking imports \& forward references. \\
sphinx-doc/sphinx & The Sphinx documentation generator. \\
UCSBarchlab/PyRTL & A collection of classes providing simple hardware specification, simulation, tracing, and testing suitable for teaching and research. \\
\addlinespace
\multicolumn{2}{c}{\textit{Web/API}} \\
\midrule
abhinavsingh/proxy.py & Ngrok FRP Alternative Fast Lightweight 0 Dependency Pluggable TLS interception DNS-over-HTTPS Poor Man's VPN Reverse \& Forward "Proxy Server" framework. \\
acesseonline/pyreportjasper & Python Reporting with JasperReports. \\
assertpy/assertpy & Simple assertion library for unit testing in python with a fluent API. \\
authlib/authlib & The ultimate Python library in building OAuth, OpenID Connect clients and servers. \\
beancount/fava & Fava - web interface for Beancount. \\
boto/boto & For the latest version of boto, Python interface to Amazon Web Services. \\
bottlepy/bottle & bottle.py is a fast and simple micro-framework for python web-applications. \\
chrismattmann/tika-python & Tika-Python is a Python binding to the Apache Tika REST services allowing Tika to be called natively in the Python community. \\
coin-or/pulp & A python Linear Programming API. \\
da4089/simplefix & Simple FIX protocol implementation for Python. \\
datnguye/dbterd & Generate the ERD as a code from dbt artifacts. \\
Doist/todoist-api-python & A python wrapper for the Todoist REST API. \\
dropbox/stone & The Official API Spec Language for Dropbox API V2. \\
falconry/falcon & The no-magic web API and microservices framework for Python developers, with a focus on reliability and performance at scale. \\
flexxui/flexx & Write desktop and web apps in pure Python. \\
jarun/buku & :bookmark: Personal mini-web in text. \\
Knio/dominate & Dominate is a Python library for creating and manipulating HTML documents using an elegant DOM API. \\
laughingman7743/PyAthena & PyAthena is a Python DB API 2.0 (PEP 249) client for Amazon Athena. \\
lepture/authlib & The ultimate Python library in building OAuth, OpenID Connect clients and servers. \\
nficano/pytube & A lightweight, dependency-free Python library (and command-line utility) for downloading YouTube Videos. \\
pallets/werkzeug & The comprehensive WSGI web application library. \\
python-twitter-tools/twitter & Python Twitter API. \\
pytube/pytube & A lightweight, dependency-free Python library (and command-line utility) for downloading YouTube Videos. \\
quantumlib/cirq & Python framework for creating, editing, and running Noisy Intermediate-Scale Quantum (NISQ) circuits. \\
rocioar/flake8-django & A flake8 plugin to detect bad practices on Django projects. \\
slackapi/python-slack-sdk & Slack Developer Kit for Python. \\
slank/awsgi & A WSGI gateway for the AWS API Gateway/Lambda proxy integration. \\
snok/django-swagger-tester & Test utility for validating OpenAPI documentation. \\
snok/drf-openapi-tester & Test utility for validating OpenAPI documentation. \\
sphinx-doc/sphinx-autobuild & Watch a Sphinx directory and rebuild the documentation when a change is detected. \\
tmux-python/libtmux & Python API / wrapper for tmux. \\
tornadoweb/tornado & Tornado is a Python web framework and asynchronous networking library, originally developed at FriendFeed. \\
twilio/twilio-python & A Python module for communicating with the Twilio API and generating TwiML. \\
\addlinespace
\multicolumn{2}{c}{\textit{ML/AI}} \\
\midrule
adbar/simplemma & Simple multilingual lemmatizer for Python, especially useful for speed and efficiency. \\
asreview/asreview & Active learning for systematic reviews. \\
eyurtsev/kor & LLM( ). \\
gunthercox/ChatterBot & ChatterBot is a machine learning, conversational dialog engine for creating chat bots. \\
pykale/pykale & Knowledge-Aware machine LEarning (KALE): accessible machine learning from multiple sources for interdisciplinary research, part of the PyTorch ecosystem. \\
PyThaiNLP/pythainlp & Thai natural language processing in Python. \\
rasbt/mlxtend & A library of extension and helper modules for Python's data analysis and machine learning libraries. \\
sorgerlab/indra & INDRA (Integrated Network and Dynamical Reasoning Assembler) is an automated model assembly system interfacing with NLP systems and databases to collect knowledge, and through a process of assembly... \\
\addlinespace
\multicolumn{2}{c}{\textit{Data/Scientific}} \\
\midrule
chezou/tabula-py & Simple wrapper of tabula-java: extract table from PDF into pandas DataFrame. \\
tqdm/tqdm & :zap: A Fast, Extensible Progress Bar for Python and CLI. \\
\addlinespace
\multicolumn{2}{c}{\textit{DevOps/Infra}} \\
\midrule
aiven/pghoard & PostgreSQL backup and restore service. \\
Amaindex/asyncio-socks-server & A SOCKS proxy server implemented with the powerful python cooperative concurrency framework asyncio. \\
amoffat/sh & Python process launching. \\
axem-solutions/dem & Containerized Development Environment Manager for embedded development. \\
cloudmarker/cloudmarker & Cloud security monitoring tool and framework. \\
Ericsson/CodeChecker & CodeChecker is an analyzer tooling, defect database and viewer extension for static and dynamic analyzer tools. \\
greenbone/autohooks & Library for managing git hooks. \\
keylime/keylime & A CNCF Project to Bootstrap \& Maintain Trust on the Edge / Cloud and IoT. \\
localstack/localstack & A fully functional local AWS cloud stack. \\
netaddr/netaddr & A network address manipulation library for Python. \\
\addlinespace
\multicolumn{2}{c}{\textit{Automation/Apps}} \\
\midrule
basnijholt/miflora & Mi Flora Plant sensor Python package. \\
cityjson/cjio & CityJSON/io: Python CLI to process and manipulate CityJSON files. \\
cylc/cylc-flow & Cylc: a general purpose workflow engine with a gift for cycling. \\
dbcli/athenacli & AthenaCLI is a CLI tool for AWS Athena service that can do auto-completion and syntax highlighting. \\
gpt-engineer-org/gpt-engineer & CLI platform to experiment with codegen. \\
home-assistant/netdisco & :mag\_right: Python library to scan local network for services and devices. \\
iterativ/openopc2 & OPC DA Python Library with awesome CLI. \\
iterative/shtab & Automagic shell tab completion for Python CLI applications. \\
juanbindez/pytubefix & Python3 library for downloading YouTube Videos. \\
ottowayi/pycomm3 & A Python Ethernet/IP library for communicating with Allen-Bradley PLCs. \\
pavdmyt/yaspin & A lightweight terminal spinner for Python with safe pipes and redirects . \\
rg3/youtube-dl & Command-line program to download videos from YouTube.com and other video sites. \\
rubik/radon & Various code metrics for Python code. \\
yt-dlp/yt-dlp & A feature-rich command-line audio/video downloader. \\
\addlinespace
\multicolumn{2}{c}{\textit{Language/Core}} \\
\midrule
23andMe/Yamale & A schema and validator for YAML. \\
alastair/python-musicbrainzngs & Python bindings for Musicbrainz' NGS webservice. \\
andymccurdy/redis-py & Redis Python client. \\
ansys/pyaedt & AEDT Python Client Package. \\
aparrish/pytracery & Python port of Kate Compton's Tracery text expansion library. \\
attwad/python-osc & Open Sound Control server and client in pure python. \\
authlib/joserfc & Implementations of JOSE RFCs in Python. \\
awslabs/aws-lambda-builders & Python library to compile, build \& package AWS Lambda functions for several runtimes \& framework. \\
beartype/beartype & Unbearably fast near-real-time pure-Python runtime-static type-checker. \\
benjaminp/six & Python 2 and 3 compatibility library. \\
berkerpeksag/astor & Python AST read/write. \\
biopython/biopython & Official git repository for Biopython (originally converted from CVS). \\
bitcraft/PyTMX & Python library to read Tiled Map Editor's TMX maps. \\
brazilian-utils/brutils & Biblioteca de utilit rios projetada para validar, gerar e manipular dados de acordo com as particularidades do Brasil . \\
BritishGeologicalSurvey/etlhelper & ETL Helper is a Python ETL library to simplify data transfer into and out of databases. \\
byroot/pysrt & Python parser for SubRip (srt) files. \\
carpedm20/emoji & emoji terminal output for Python. \\
celery/billiard & Multiprocessing Pool Extensions. \\
celery/celery & Distributed Task Queue (development branch). \\
celery/kombu & Messaging library for Python. \\
CheetahTemplate3/cheetah3 & Cheetah3 is a free (MIT) and open source template engine for Python. \\
cherrypy/cherrypy & CherryPy is a pythonic, object-oriented HTTP framework. \\
chrthomsen/pygrametl & Official repository for pygrametl - ETL programming in Python. \\
cjrh/aiorun & A "run" function for asyncio-based apps that does all the boilerplate. \\
construct/construct & Construct: Declarative data structures for python that allow symmetric parsing and building. \\
cool-RR/PySnooper & Never use print for debugging again. \\
csingley/ofxtools & Python OFX Library. \\
ctypesgen/ctypesgen & Pure-python wrapper generator for ctypes. \\
cython/cython & The most widely used Python to C compiler. \\
danijar/handout & Turn Python scripts into handouts with Markdown and figures. \\
darcymason/pydicom & Read, modify and write DICOM files with python code. \\
davidhalter/jedi & Awesome autocompletion, static analysis and refactoring library for python. \\
davidhalter/parso & A Python Parser. \\
davidlatwe/montydb & Monty, Mongo tinified. \\
defnull/multipart & A fast multipart/form-data parser for python. \\
Delgan/loguru & Python logging made (stupidly) simple. \\
didix21/mdutils & Python package contains a set of basic tools that can help to create a markdown file. \\
djc/couchdb-python & Python library for working with CouchDB. \\
dulwich/dulwich & Pure-Python Git implementation. \\
enkore/i3pystatus & A complete replacement for i3status. \\
erikrose/parsimonious & The fastest pure-Python PEG parser I can muster. \\
fabiocaccamo/python-fsutil & :computer: :wrench: high-level file-system operations for lazy devs. \\
facebook/TestSlide & A Python test framework. \\
facebookincubator/TestSlide & A Python test framework. \\
facelessuser/coloraide & A library to aid in using colors. \\
fastavro/fastavro & Fast Avro for Python. \\
geomet/geomet & GeoMet - Pure Python conversion library for common geospatial data formats. \\
geopy/geopy & Geocoding library for Python. \\
GiacomoPope/dilithium-py & A pure python implementation of ML-DSA (FIPS 204) and CRYSTALS-Dilithium. \\
GiacomoPope/kyber-py & A pure python implementation of ML-KEM (FIPS 203) and CRYSTALS-Kyber. \\
globocom/m3u8 & Python m3u8 Parser for HTTP Live Streaming (HLS) Transmissions. \\
google/latexify\_py & A library to generate LaTeX expression from Python code. \\
gpiozero/gpiozero & A simple interface to GPIO devices with Raspberry Pi. \\
gunyarakun/python-shogi & A pure Python shogi library with move generation and validation and handling of common formats. \\
h2non/filetype.py & Small, dependency-free, fast Python package to infer binary file types checking the magic numbers signature. \\
hardbyte/python-can & The can package provides controller area network support for Python developers. \\
heuer/segno & Python QR Code and Micro QR Code encoder. \\
hgrecco/pint & Operate and manipulate physical quantities in Python. \\
ikalchev/HAP-python & A python implementation of the HomeKit Accessory Protocol (HAP). \\
Instagram/LibCST & A concrete syntax tree parser and serializer library for Python that preserves many aspects of Python's abstract syntax tree. \\
izimobil/polib & Pure python library to manipulate, create, modify gettext files. \\
jacobschaer/python-doipclient & Pure Python ISO 13400 Client. \\
jamesoff/simplemonitor & A Python-based network and host monitor. \\
jazzband/geojson & Python bindings and utilities for GeoJSON. \\
jd/tenacity & Retrying library for Python. \\
johnpaulett/python-hl7 & A simple library for parsing messages of Health Level 7 version 2.x into Python objects. \\
joowani/binarytree & Python Library for Studying Binary Trees. \\
josegonzalez/python-github-backup & backup a github user or organization. \\
kanaka/websockify & Websockify is a WebSocket to TCP proxy/bridge. \\
karlicoss/orgparse & Python module for reading Emacs org-mode files. \\
kennethreitz/envoy & Python Subprocesses for Humans . \\
kislyuk/argcomplete & Python and tab completion, better together. \\
kivy/buildozer & Generic Python packager for Android and iOS. \\
kivy/kivy & Open source UI framework written in Python, running on Windows, Linux, macOS, Android and iOS. \\
kivy/oscpy & An efficient OSC implementation compatible with python2.7 and 3.5+. \\
lark-parser/lark & Lark is a parsing toolkit for Python, built with a focus on ergonomics, performance and modularity. \\
Lawouach/WebSocket-for-Python & WebSocket client and server library for Python 2 and 3 as well as PyPy. \\
lepture/mistune & A fast yet powerful Python Markdown parser with renderers and plugins. \\
lmc2179/bayesian\_bootstrap & bayesian bootstrapping in python. \\
m-labs/migen & A Python toolbox for building complex digital hardware. \\
mahmoud/boltons & Like builtins, but boltons. \\
majiidd/persiantools & Jalali date and datetime with other tools. \\
mapado/haversine & Calculate the distance between 2 points on Earth. \\
marcelotduarte/cx\_Freeze & Creates standalone executables from Python scripts with the same performance as the original script. \\
martinblech/xmltodict & Python module that makes working with XML feel like you are working with JSON. \\
masoniteframework/orm & Masonite ORM is a beautiful Python ORM. \\
maxmind/GeoIP2-python & Python code for GeoIP2 webservice client and database reader. \\
maxmind/MaxMind-DB-Reader-python & Python MaxMind DB reader extension. \\
mementum/backtrader & Python Backtesting library for trading strategies. \\
metabrainz/picard & Picard is a cross-platform music tagger powered by the MusicBrainz database. \\
mewwts/addict & The Python Dict that's better than heroin. \\
Microsoft/ptvsd & Python debugger package for use with Visual Studio and Visual Studio Code. \\
mido/mido & MIDI Objects for Python. \\
mirumee/google-i18n-address & Google's i18n address data packaged for Python. \\
missionpinball/mpf & Mission Pinball Framework: Open source software to run a real pinball machine. \\
mitmproxy/mitmproxy & An interactive TLS-capable intercepting HTTP proxy for penetration testers and software developers. \\
miyuchina/mistletoe & A fast, extensible and spec-compliant Markdown parser in pure Python. \\
mnaberez/py65 & Emulate 6502-based microcomputer systems in Python. \\
mopidy/mopidy & Mopidy is an extensible music server written in Python. \\
moskytw/mosql & Build SQL with native Python data structure smoothly. \\
msiemens/tinydb & TinyDB is a lightweight document oriented database optimized for your happiness. \\
mu-editor/mu & A small, simple editor for beginner Python programmers. \\
nackjicholson/aiosql & Simple SQL in Python. \\
ncclient/ncclient & Python library for NETCONF clients. \\
neogeny/TatSu & TatSu generates Python parsers from grammars in a variation of EBNF. \\
Neoteroi/rodi & Implementation of dependency injection for Python 3. \\
nickreynke/python-gedcom & Python module for parsing, analyzing, and manipulating GEDCOM files. \\
nicotine-plus/nicotine-plus & Graphical client for the Soulseek peer-to-peer network. \\
nioinnovation/python-xbee & Python tools for working with XBee radios. \\
noahmorrison/chevron & A Python implementation of mustache. \\
ntoll/uflash & A module and command to easily flash Python onto the BBC's micro:bit device. \\
petertodd/python-bitcoinlib & Python3 library providing an easy interface to the Bitcoin data structures and protocol. \\
pex-tool/pex & A tool for generating .pex (Python EXecutable) files, lock files and venvs. \\
pexpect/pexpect & A Python module for controlling interactive programs in a pseudo-terminal. \\
piqueserver/piqueserver & An Ace of Spades 0.75 server based on PySnip. \\
piskvorky/smart\_open & Utils for streaming large files (S3, HDFS, gzip, bz2...). \\
plasma-umass/scalene & Scalene: a high-performance, high-precision CPU, GPU, and memory profiler for Python with AI-powered optimization proposals. \\
pmaupin/pdfrw & pdfrw is a pure Python library that reads and writes PDFs. \\
pointhi/kicad-footprint-generator & creating kicad footprints using python scripts. \\
pre-commit/pre-commit & A framework for managing and maintaining multi-language pre-commit hooks. \\
projectmesa/mesa & Mesa is an open-source Python library for agent-based modeling, ideal for simulating complex systems and exploring emergent behaviors. \\
psf/requests & A simple, yet elegant, HTTP library. \\
py-moneyed/py-moneyed & Provides Currency and Money classes for use in your Python code. \\
PyCQA/autoflake & Removes unused imports and unused variables as reported by pyflakes. \\
pycqa/eradicate & Removes commented-out code from Python files. \\
pydantic/pydantic & Data validation using Python type hints. \\
pydcs/dcs & Digital Combat Simulator Python mission framework. \\
pydicom/pydicom & Read, modify and write DICOM files with python code. \\
pylessard/python-udsoncan & Python implementation of UDS (ISO-14229) standard. \\
pymedphys/pymedphys & A community effort to develop an open standard library for Medical Physics in Python. \\
pymodbus-dev/pymodbus & A full modbus protocol written in python. \\
pympler/pympler & Development tool to measure, monitor and analyze the memory behavior of Python objects in a running Python application. \\
pypa/setuptools-scm & the blessed package to manage your versions by scm tags. \\
pyparsing/pyparsing & Python library for creating PEG parsers. \\
pyserial/pyserial & Python serial port access library. \\
pysimplesoap/pysimplesoap & Python Simple SOAP Library. \\
python-jsonschema/jsonschema & An implementation of the JSON Schema specification for Python. \\
Python-Markdown/markdown & A Python implementation of John Gruber s Markdown with Extension support. \\
python-mechanize/mechanize & The official source code for the python-mechanize project. \\
python-poetry/poetry & Python packaging and dependency management made easy. \\
python-telegram-bot/python-telegram-bot & We have made you a wrapper you can't refuse. \\
pyvisa/pyvisa & A Python package with bindings to the "Virtual Instrument Software Architecture" VISA library, in order to control measurement devices and test equipment via GPIB, RS232, or USB. \\
qtile/qtile & :cookie: A full-featured, hackable tiling window manager written and configured in Python (X11 + Wayland). \\
quodlibet/mutagen & Python module for handling audio metadata. \\
r1chardj0n3s/parse & Parse strings using a specification based on the Python format() syntax. \\
realpython/codetiming & A flexible, customizable timer for your Python code. \\
reddit/baseplate.py & reddit's python service framework. \\
rhinstaller/dasbus & DBus library in Python 3. \\
richardkiss/pycoin & Python-based Bitcoin and alt-coin utility library. \\
riptideio/pymodbus & A full modbus protocol written in python. \\
rm-hull/luma.core & A component library providing a Pillow-compatible drawing canvas, and other functionality to support drawing primitives and text-rendering capabilities for small displays on the Raspberry Pi and ot... \\
rm-hull/OPi.GPIO & RPi.GPIO drop-in replacement library for Orange Pi Zero and other SBCs. \\
robotpy/cxxheaderparser & Modern pure python C++ header parser. \\
roniemartinez/latex2mathml & Pure Python library for LaTeX to MathML conversion. \\
rossengeorgiev/aprs-python & Python module for working with APRS. \\
rr-/docstring\_parser & Parse Python docstrings in various flavors. \\
rsalmei/clearly & Clearly see and debug your celery cluster in real time! \\
rtfd/commonmark.py & DEPRECATED: Python CommonMark parser. \\
rthalley/dnspython & a powerful DNS toolkit for python. \\
sartography/SpiffWorkflow & A powerful workflow engine implemented in pure Python. \\
savoirfairelinux/num2words & Modules to convert numbers to words. \\
sdispater/poetry & Python packaging and dependency management made easy. \\
sibson/vncdotool & A command line VNC client and python library. \\
SolarEdgeTech/pyctuator & Monitor Python applications using Spring Boot Admin. \\
sopel-irc/sopel & :robot::speech\_balloon: An easy-to-use and highly extensible IRC Bot framework. \\
sosy-lab/benchexec & BenchExec: A Framework for Reliable Benchmarking and Resource Measurement. \\
spyder-ide/spyder & Official repository for Spyder - The Scientific Python Development Environment. \\
ssato/python-anyconfig & Python library provides common APIs to load and dump configuration files in various formats. \\
Tabviewer/tabview & Python curses command line CSV and tabular data viewer. \\
tariqdaouda/pyArango & Python Driver for ArangoDB with built-in validation. \\
terrencepreilly/darglint & A python documentation linter which checks that the docstring description matches the definition. \\
textX/Arpeggio & Parser interpreter based on PEG grammars. \\
tkarabela/pysubs2 & A Python library for editing subtitle files. \\
tna0y/Python-random-module-cracker & Predict python's random module generated values. \\
tomerfiliba-org/rpyc & RPyC (Remote Python Call) - A transparent and symmetric RPC library for python. \\
tryexceptpass/sofi & an OS agnostic UI module for Python. \\
Turbo87/utm & Bidirectional UTM-WGS84 converter for python. \\
tweag/FawltyDeps & Python dependency checker. \\
ukBaz/python-bluezero & A simple Python interface to Bluez. \\
urwid/urwid & Console user interface library for Python (official repo). \\
vzhd1701/evernote-backup & Backup \& export all Evernote notes and notebooks. \\
wbolster/jsonlines & python library to simplify working with jsonlines and ndjson data. \\
westerndigitalcorporation/pyvcd & Python package for writing Value Change Dump (VCD) files. \\
wolverdude/genson & GenSON is a powerful, user-friendly JSON Schema generator built in Python. \\
Yubico/python-fido2 & Provides library functionality for FIDO 2.0, including communication with a device over USB. \\
yuma-m/pychord & Python library to handle musical chords. \\
\addlinespace
\multicolumn{2}{c}{\textit{Other}} \\
\midrule
afewmail/afew & an initial tagging script for notmuch mail. \\
akaihola/pgtricks & Handy helpers for PostgreSQL users. \\
aparrish/pronouncingpy & A simple interface for the CMU pronouncing dictionary. \\
bbangert/beaker & WSGI middleware for sessions and caching. \\
bgreenlee/pygtail & Pygtail reads log file lines that have not been read. \\
btclib-org/btclib & btclib: a python3 library for 'bitcoin cryptography'. \\
cdump/investments & Analysis of Interactive Brokers reports for tax reporting. \\
CenterForOpenScience/pydocx & An extendable docx file format parser and converter. \\
Ch00k/ffmpy & Pythonic interface for FFmpeg/FFprobe command line. \\
chipsec/chipsec & Platform Security Assessment Framework. \\
citrusvanilla/tinyflux & The tiny time series database optimized for your happiness. \\
conda/conda-pack & Package conda environments for redistribution. \\
daily-co/dailyai & Open Source framework for voice and multimodal conversational AI. \\
elastic/rally & Macrobenchmarking framework for Elasticsearch. \\
exo-lang/exo & Exocompilation for productive programming of hardware accelerators. \\
fetchai/agents-aea & A framework for autonomous economic agent (AEA) development. \\
GNS3/gns3-server & GNS3 server. \\
GoogleCloudPlatform/snapshot-debugger & A software repository. \\
jaraco/inflect & Correctly generate plurals, ordinals, indefinite articles; convert numbers to words. \\
jeremiah-c-leary/vhdl-style-guide & Style guide enforcement for VHDL. \\
kootenpv/access\_points & Scan your WiFi and get access point information and signal quality. \\
markdrago/pgsanity & Check syntax of postgresql sql files. \\
mciepluc/cocotb-coverage & Functional Coverage and Constrained Randomization Extensions for Cocotb. \\
meerk40t/meerk40t & Hackable Laser software for K40 / GRBL / Fibre Lasers. \\
meerk40t/svgelements & SVG Parsing for Elements, Paths, and other SVG Objects. \\
Michael-F-Bryan/auto-changelog & A small program that will generate a changelog from git repos using "conventional style" commit messages. \\
MycroftAI/adapt & Adapt Intent Parser. \\
myint/cppclean & Finds problems in C++ source that slow development of large code bases. \\
mystor/git-revise & A handy tool for doing efficient in-memory commit rebases \& fixups. \\
NewFuture/DDNS & IP( dnspod, DNS,CloudFlare, ,DNSCOM...). \\
opensearch-project/OpenSearch-Benchmark & OpenSearch Benchmark - a community driven, open source project to run performance tests for OpenSearch. \\
openSUSE/osc & The Command Line Interface to work with an Open Build Service. \\
OSInside/kiwi & KIWI - Appliance Builder Next Generation. \\
pipecat-ai/pipecat & Open Source framework for voice and multimodal conversational AI. \\
pycontribs/ansi2html & Convert text with ansi color codes to HTML. \\
pypa/auditwheel & Auditing and relabeling cross-distribution Linux wheels. \\
pypa/setuptools & Official project repository for the Setuptools build system. \\
pypiserver/pypiserver & Minimal PyPI server for uploading \& downloading packages with pip/easy\_install. \\
rasterio/affine & Affine transformation matrices. \\
snapcore/snapcraft & Package, distribute, and update any app for Linux and IoT. \\
Snawoot/postfix-mta-sts-resolver & Daemon which provides TLS client policy for Postfix via socketmap, according to domain MTA-STS policy. \\
summa-tx/riemann & rapid prototyping transaction toolbox for Bitcoin-style chains . \\
Supervisor/supervisor & Supervisor process control system for Unix (supervisord). \\
thammegowda/mtdata & A tool that locates, downloads, and extracts machine translation corpora. \\
tobi-wan-kenobi/bumblebee-status & bumblebee-status is a modular, theme-able status line generator for the i3 window manager. \\
tomerfiliba/plumbum & Plumbum: Shell Combinators. \\
vd2org/snowflake & The Snowflake generator done right. \\
vivisect/vivisect & A repository related to reverse-engineering, emulation. \\
\addlinespace
\end{longtable}